\documentclass[aps,prc,preprint,groupedaddress,showpacs,floatfix,nofootinbib]{revtex4}
\usepackage{amssymb}
\usepackage{graphicx}
\usepackage{bm}

\newcommand{\be}{\begin{equation}}
\newcommand{\ee}{\end{equation}}
\newcommand{\bel}[1]{\be\label{#1}}
\newcommand{\re}[1]{Eq.~(\ref{#1})}
\newcommand{\ds}{\displaystyle}
\newcommand{\ov}[1]{\overline{#1}}
\newcommand{\hsp}{\hspace*{1pt}}
\newcommand{\hspm}{\hspace*{0.5pt}}

\begin{document}

\title{Evolution of antibaryon abundances\\
in the early Universe and in heavy-ion collisions}

\author{L.M. Satarov}
\affiliation{Frankfurt Institute for Advanced Studies,~D--60438 Frankfurt am Main, Germany}
\affiliation{National Research Center ''Kurchatov Institute'', 123182 Moscow, Russia}

\author{I.N.~Mishustin}

\affiliation{Frankfurt Institute for Advanced Studies,~D--60438 Frankfurt am Main, Germany}
\affiliation{National Research Center ''Kurchatov Institute'', 123182 Moscow, Russia}

\author{W. Greiner}

\affiliation{Frankfurt Institute for Advanced Studies,~D--60438 Frankfurt am Main, Germany}

\begin{abstract}
We study the kinetics of antibaryon production and annihilation in an expanding system, assuming that it is spatially
homogeneous and chemically equilibrated at the initial stage. By solving simplified rate equations for (anti)baryon abundances we study the deviations from chemical equilibrium at late stages. The calculations are done for different expansion rates and net-baryon-to-entropy ratios, covering the conditions from early Universe to heavy--ion collisions. Our analysis includes both stable (anti)baryons and resonances.
We conclude that residual antibaryon abundances are very sensitive to the time scales of expansion. Our calculations naturally explain noticeable deviations of $\ov{p}/\pi$ and $p/\pi$ ratios observed in nuclear collisions at the LHC energy from the thermal model predictions.
We conclude that at high bombarding energies the chemical freeze-out of (anti)baryons should occur at lower temperatures as compared to mesons.
\end{abstract}


\maketitle

\section{Introduction}
Mechanisms of (anti)baryon production and annihilation are still not well understood both microscopically (in hadronic and heavy--ion
reactions) as well as at a global level (in the early Universe). It is argued~\cite{Koc86} that enhanced yields of multi-strange
antibaryons in heavy--ion collisions can be considered as a signature of the deconfined phase of strongly-interacting matter, the quark-gluon plasma (QGP). Some collective mechanisms which may result in enhanced production of
antibaryons and antinuclei in nuclear collisions have been discussed in Refs.~\cite{Sch91,Sor92,Mis93,Mis95,Ble00,Mis05,Lar10}. Possible reasons for suppressing antibaryon annihilation in dense hadronic matter have been suggested in~\cite{Mis05,Spi96,Pan97}.
Increased interest to the issue of antimatter production has been stimulated recently by observations~\cite{Abe10,Aga11} of anti(hyper)nuclei in Au+Au collisions at the RHIC bombarding energy $\sqrt{s_{\rm NN}}=200~\textrm{GeV}$.

Traditional cascade models based on binary hadronic interactions and vacuum cross sections fail to reproduce antibaryon yields observed in relativistic heavy--ion collisions. For example, the UrQMD calculations~\cite{Ble00} underestimate experimental antiproton multiplicities in central Pb+Pb collisions at the SPS bombarding energy $E_{\hsp\rm lab}=158~\textrm{A\hsp GeV}$ by factor of about 3. Even larger discrepancies with observed antiproton yields have been obtained~\cite{Spi96} for lower AGS energies. As proposed
in~Refs.~\cite{Ble00,Rap01,Cgr01}, this disagreement may be explained by
nonbinary multihadron interactions which are disregarded in conventional transport simulations. The direct calculations within the HSD model showed~\cite{Cas02} that such interactions, indeed, give important contributions to (anti)baryon production at AGS and SPS energies. Effectively, these $B\ov{B}$ production channels compensate to a large extent losses of antibaryons due to their annihilation.
It is natural to assume that they should be even more important at higher RHIC and LHC energies.

On the other hand, estimates of hadron yields obtained within thermal models~\mbox{\cite{And06,And09,Cle06}} agree rather well with experimental data on heavy--ion collisions in a broad range of bombarding energies and centralities. As demonstrated in Refs.~\cite{And11,Cle12}, even yields
of composite antinuclei can be well reproduced in such an approach. The latter assumes  that hadrons are produced at the decay of thermally and chemically equilibrated source, a ''fireball''. The temperature and chemical potential of the fireball are considered to be functions of the bombarding energy. They are determined from the best fit of hadron ratios observed in nuclear collisions at various energies. It is believed that hadron multiplicities do not change noticeably during the subsequent expansion and cooling of the fireball until its kinetic freeze-out.

Recent data of ALICE Collaboration~\cite{Abe12} reveal significant deviations from predictions of the thermal model. The latter overestimates the $p/\pi^+$ and $\ov{p}/\pi^-$ ratios observed in central Pb+Pb collisions at $\sqrt{s_{\rm NN}}=2.76~\textrm{TeV}$ by a factor of
approximately 1.5. On the other hand, relative yields of $\pi^\pm$ and $K^\pm$ mesons in the same reaction are well described.
The authors of Ref.~\cite{Ste13} analyzed the ALICE data within a hydro-cascade approach and found that due to
annihilation at late stages of the reaction the (anti)baryon yields should be significantly reduced. Note, however, that
the $B\ov{B}$ production in multihadron interactions was not included explicitly in these calculations. It was found in Ref.~\cite{Bec12}
that exclusion of antibaryons from the thermal fit of data observed at the SPS energies increases the effective temperature
for other hadrons. This gives an evidence in favor of separate freeze-out stages for mesons and antibaryons. The possibility of later
freeze-out of (anti)baryons in heavy--ion collisions at RHIC and LHC energies was also discussed in Ref.~\cite{Bas00}.

In relation to this problem, we would like to mention theoretical studies of residual antibaryon abundances in the early Universe. The authors of Refs.~\cite{Zel71,Sch86,Kol90} solve the rate equations for $N,\ov{N}$ abundances, which includes both the annihilation
and production terms. The production rates were estimated by using the detailed balance arguments (see below). It has been shown that even small initial baryon asymmetries lead to extremely small $\ov{N}/N$ ratios at the current stage of the Universe evolution.
Later on similar rate equations were used in Refs.~\cite{Bir83,Cgr01,Sch09,Nor10,Pan12} to investigate the (anti)baryon production and annihilation in heavy--ion collisions.

In this paper we use essentially the same formalism to study evolution of (anti)baryon abundance both in the early Universe
and in nuclear collisions. As compared to previous models, we take into account not only (anti)nucleons but also
heavier (anti)baryonic species. We apply our equation of state (EoS) with excluded volume corrections~\cite{Sat09}
to calculate equilibrium hadronic densities which are used to determine the $B\ov{B}$ production terms.
All important hadronic species are taken into account to calculate the expansion rate of cosmic matter as a function of temperature. Our predictions for (anti)baryon yields in heavy--ion collisions are compared with data obtained at the SPS, RHIC and LHC
energies.

The paper is organized as follows. In Sec.~II we develop simplified rate equations to describe the evolution of (anti)baryon
abundances in an expanding hadronic system. Section~III is devoted to evolution of (anti)baryon abundances in the early Universe.
In Sec.~IV this formalism is applied for relativistic heavy--ion collisions and detailed comparison with existing experimental data is made. The summary and outlook are given in Sec.~V.

\section{Rate equations for (anti)baryons\label{2-1}}
Let us consider a hadronic system consisting of mesons ($M=\pi,K,\ov{K},\rho\ldots $),
baryons ($B=N,\Lambda,\Sigma,\Delta\ldots $) and corresponding antibaryons. We assume that this system
is at thermal (but not necessarily in chemical) equilibrium at some temperature $T$.
The evolution of antibaryon abundances is determined mainly by competition of two hadronic processes, namely, the annihilation ($B\ov{B}\to M_1M_2\ldots$) and production (e.g. $M_1M_2\ldots\to B\ov{B}$) reactions.
Let $\sigma_{\ov{i}k}^{\rm\hsp ann}$ to denote the annihilation cross section of the $i$--th antibaryon interaction with $k$--th baryons. This cross section is a function of the relative velocity $v_{\ov{i}k}$\hsp. The annihilation loss of the $i$--th antibaryons (per unit time and volume) can be written as\hsp\footnote
{
\hsp Multiparticle annihilation processes~\cite{Mis05} are disregarded here.
}
\bel{rann}
\left(\frac{\ds d^{\hsp 4} N_{\ov{i}}}{\ds d^{\hsp 4}x}\right)_{\rm ann}=-\sum\limits_{k}<\sigma^{\rm ann}_{\ov{i}k}v_{\hsp\ov{i}k}>
n_{\hsp\ov{i}}\hsp\hsp n_k\hsp,
\ee
where $n_{\ov{i}}$ and $n_{k}$ are the partial densities of $i$-th antibaryons and $k$-th baryons, respectively.
The sum in \re{rann} runs over all stable ($k=N,\Lambda,\Sigma,\Xi,\Omega$) and unstable ($k=\Delta,N^*,\Lambda^*\ldots$)
baryonic species. The angular brackets denote averaging over thermal distributions of corresponding (anti)baryons. In the following we
neglect possible in-medium modifications of annihilation cross sections (see, however, Ref.~\cite{Mis05}).

The rate of $i$-th antibaryon production in inverse reactions $M_1M_2\ldots\to\ov{i}k$ can be estimated by using the detailed
balance principle. In particular, multimeson inelastic interactions which include more than two mesons in the initial state should be rather important in dense hadronic matter~\cite{Rap01,Cgr01,Cas02}\hsp\footnote
{
The neglect of such interactions leads to violation of the detailed balance and underestimation of antibaryon abundances~\cite{Ble00} in most existing transport models of heavy-ion collisions.
}.
For small deviations from chemical equilibrium, an approximate expression for the production rate
can be obtained from the r.h.s. of~\re{rann} with the replacement
\mbox{$n_{\hsp\ov{i}}\hsp\hsp n_k\to -n_{\hsp\ov{i}}^{(\rm eq)}\hsp\hsp n_k^{(\rm eq)}$}, where $n_{\ov{i}}^{(\rm eq)}$ and $n_k^{(\rm eq)}$ are the corresponding equilibrium densities. The resulting rate equation can be written as follows
\bel{netr}
\frac{\ds d^{\hsp 4} N_{\ov{i}}}{\ds d^{\hsp 4}x}=\sum\limits_{k}<\sigma^{\rm ann}_{\ov{i}k}v_{\hsp\ov{i}k}>
\left[n_{\hsp\ov{i}}^{(\rm eq)}\hsp\hsp n_k^{(\rm eq)}-n_{\hsp\ov{i}}\hsp\hsp n_k\right].
\vspace*{-2mm}
\ee
In the case of unstable antibaryons with a nonzero width $\Gamma_{i}$, one should also add the term
$\Gamma_{i}\hsp\hsp (n_{\ov{i}}^{(\rm eq)}-n_{\ov{i}})$ to the r.h.s. of~\re{netr} \mbox{\cite{Bir83,Ko88}}.
It describes the decay of $i$-th antibaryons into lighter states as well as their regeneration in antibaryon-meson collisions. We do not include such terms explicitly, because they do not change the total antibaryon density $n_{\ov{B}}=\sum\limits_{i}n_{\ov{i}}$.

In our qualitative analysis we neglect spatially inhomogeneous (e.g. surface) effects, assuming that all particle densities and temperature are only functions of time $t$. In this approximation, denoting by $V=V(t)$ the total volume of system\hsp\footnote
{
In the case of expanding Universe (see Sec. III)  the so-called ''comoving'' volume $V=4\pi R^3/3$ will be introduced, where $R$ is related to the Hubble parameter $H=\dot{R}/R$.
}, one has
\bel{lhsr}
\frac{\ds d^{\hsp 4} N_{\ov{i}}}{\ds d^{\hsp 4}x}= \frac{1}{V}\frac{d\hsp (n_{\hsp\ov{i}}V)}{dt}=\dot{n}_{\hsp\ov{i}}+n_{\hsp\ov{i}}\hsp\frac{\dot{V}}{V}\,.
\ee
The last term in this equation takes into account the ''trivial'' reduction of the antibaryon density due to the (uniform) system expansion.
Substituting~(\ref{lhsr}) into (\ref{netr}) gives the set of coupled rate equations for antibaryon densities. Analogous differential equations for baryon densities $n_i$ are obtained from Eqs.~(\ref{netr})--(\ref{lhsr}) after the replacements $\ov{i}\to i$ and $k\to\ov{k}$.

Up to now, the experimental information about annihilation cross sections
for antibaryons heavier than antinucleons is very scarce. As far as we know, some theoretical estimates exist only for antihyperons~\cite{Kap02}. In the following we assume that probabilities of the $\ov{i}\hsp k$ and $\ov{N}N$ annihilations are approximately equal:
$\left<\sigma^{\rm\hsp ann}_{\ov{i}\hsp k}v_{\ov{i}\hsp k}\right>\simeq\left<\sigma_{\rm\hsp ann}v_{\rm rel}\right>$, where
$v_{\hspm\rm rel}=v_{\ov{N}N}$ and $\sigma_{\rm ann}=\sigma^{\hsp\rm ann}_{\ov{N}N}$\hsp .
Using this relation in~\re{netr} and taking sum over all $\ov{i}$, one arrives at the equations:
\vspace*{2mm}
\bel{kineq1}
\frac{1}{V}\frac{\ds d\hsp (n_{\ov{B}}\hsp V)}{\ds dt}=\frac{1}{V}\frac{\ds d\hsp (n_{B}\hsp V)}{\ds dt}=
\left<\sigma_{\rm ann}v_{\hspm\rm rel}\right>\left(n_B^{(\rm eq)}\hsp n_{\ov{B}}^{(\rm eq)}-n_B\hsp n_{\ov{B}}\right),
\vspace*{2mm}
\ee
where $n_B=\sum\limits_k n_k$, and $n_B^{(\rm eq)}, n_{\ov{B}}^{(\rm eq)}$ are the equilibrium values of~$n_B, n_{\ov{B}}$\hsp .

One can formally exclude the system volume by taking into account the conservation of the total entropy $S_{\hsp\rm tot}=sV$ where $s$ is the entropy density of considered matter. Below we calculate this quantity\hsp\footnote
{
In calculating the entropy density we neglect deviations from chemical equilibrium. At given $T$ and~$\mu$ we find
the strange chemical potential $\mu_S$~\cite{Sat09} from the condition of strangeness neutrality.
}
as a function of temperature and baryon chemical potential~$\mu$ by using the EoS of ideal hadron gas with excluded volume corrections~\mbox{\cite{Ris91,Sat09}}. We choose the same excluded volume parameter $v=1~\textrm{fm}^3$ for all hadronic
species~\cite{Sat09}. Our EoS includes all known hadrons with masses up to 2 GeV in the zero width approximation\hsp\footnote
{
Following  Ref.~\cite{And09}, we take into account the contribution of $\sigma$ meson resonance with parameters \mbox{$m_\sigma=484~\textrm{MeV}$} and $\Gamma_\sigma=510$ MeV.
}.
In the case of expanding Universe we add to the entropy the contributions of photons and leptons.

The condition $S_{\hsp\rm tot}=\textrm{const}$ can be written as
\bel{entrc}
s\hsp (T,\mu)V(t)=s\hsp (T_0,\mu_0)V(t_0)\,,
\vspace*{-2mm}
\ee
where $T_0$ and $\mu_0$ are the values of temperature and baryon chemical potential at the initial time $t=t_0$.
Below, we also take into account the conservation of the net baryon number, $B_{\hsp\rm net}=n_{\hspm\rm net}\hsp V=\textrm{const}$,
where $n_{\hspm\rm net}= n_B-n_{\ov{B}}$.  One can see that this condition automatically follows from the first equality in~\re{kineq1}.
It is convenient to introduce the specific entropy
\bel{adcon}
\sigma\equiv\frac{S_{\hsp\rm tot}}{B_{\hsp\rm net}}=\frac{s}{n_{\hspm\rm net}}\,,
\ee
which does not contain the extensive variable $V$ and remains constant during the isentropic expansion. At fixed
$\sigma$, the quantities $\mu, n_{\ov{B}}^{(\rm eq)}, n_{B}^{(\rm eq)}$ can be regarded as functions of temperature.

Let us introduce the dimensionless quantity \mbox{$Y=n_{\ov{B}}/s$} which is proportional to the multiplicity of antibaryons in the volume $V$. From~\re{entrc} we get the relation \mbox{$V^{-1}d\hsp (n_{\ov{B}}\hsp V)/d\hsp t=s\dot{Y}$}\hspace*{-1pt}. Using further (\ref{adcon}), one can finally rewrite~\re{kineq1} in the form
\bel{kineqa2}
\dot{Y}=\Gamma\left[Y_{\rm eq}\hsp (Y_0+Y_{\rm eq})-Y\hsp (Y_0+Y)\right],
\ee
where $Y_0=\sigma^{-1}$ is the baryon asymmetry parameter, $\Gamma=s\left<\hspm\sigma_{\rm ann}v_{\hspm\rm rel}\right>$ and
$Y_{\rm eq}=n_{\ov{B}}^{(\rm eq)}/s$ is the value of $Y$ at chemical equilibrium. Note, that in baryon--symmetric equilibrium
matter~($Y_0=0$) the chemical potentials of hadrons vanish, $\mu=\mu_S=0$.

In the following we apply the parametrization~\cite{Dov92} ($c=\hbar=1$)
\bel{pancr}
\sigma_{\rm ann}=\left(38+\frac{35}{v_{\hspm\rm rel}}\right)\,\textrm{mb}\,.
\ee
Here and below we neglect Coulomb and isospin effects. In the nonrelativistic approximation one gets the estimate
\bel{pancr1}
<\sigma_{\rm ann}\hsp v_{\hspm\rm rel}>=\left(3.5+3.8\hsp <v_{\hspm\rm rel}>\right)\hsp\textrm{fm}^2c\simeq\left(3.5+8.6/\sqrt{x}\right)\textrm{fm}^2c\hsp ,
\ee
were $x=m_N/T$ ($m_N=939~\textrm{MeV}$ is the nucleon mass).

\section{Antibaryons in the early Universe}

Let as consider the stage of the Universe evolution, corresponding to the lepto--hadronic era~\cite{Zel71,Kol90} when most abundant particles are photons ($\gamma$), leptons ($e^\pm,\mu^\pm,\nu,\hsp\ov{\nu}$) and hadrons (mesons and baryon-antibaryon pairs).
The low temperature end of this era corresponds to the values $T\sim 1~\textrm{MeV}$. At lower temperatures the $e^+e^-$ annihilation starts and neutrinos decouple. Later on the nucleosynthesis processes become important. We assume that hadrons appear during hadronization of the QGP, when temperature drops below some critical value \mbox{$T_c\sim 17\hspm 0$~MeV}. Approximately, the lepto--hadronic era corresponds to
the time interval between 10$^{-5}$~$s$ and 1~$s$ from the Big Bang. Unless stated otherwise, we neglect possible deviations from thermal and chemical equilibrium of cosmic matter in the considered temperature interval. Having in mind small baryon asymmetry of our Universe
($Y_0\ll 1$) we calculate all thermodynamic quantities, e.g. energy- and entropy densities, in the baryon--free limit $\mu=\mu_S=0$. In this approximation, these quantities are functions of temperature only.

 \begin{figure*}[hbt!]
          \centerline{\includegraphics[trim=0 8.5cm 0 8cm, clip, width=\textwidth]{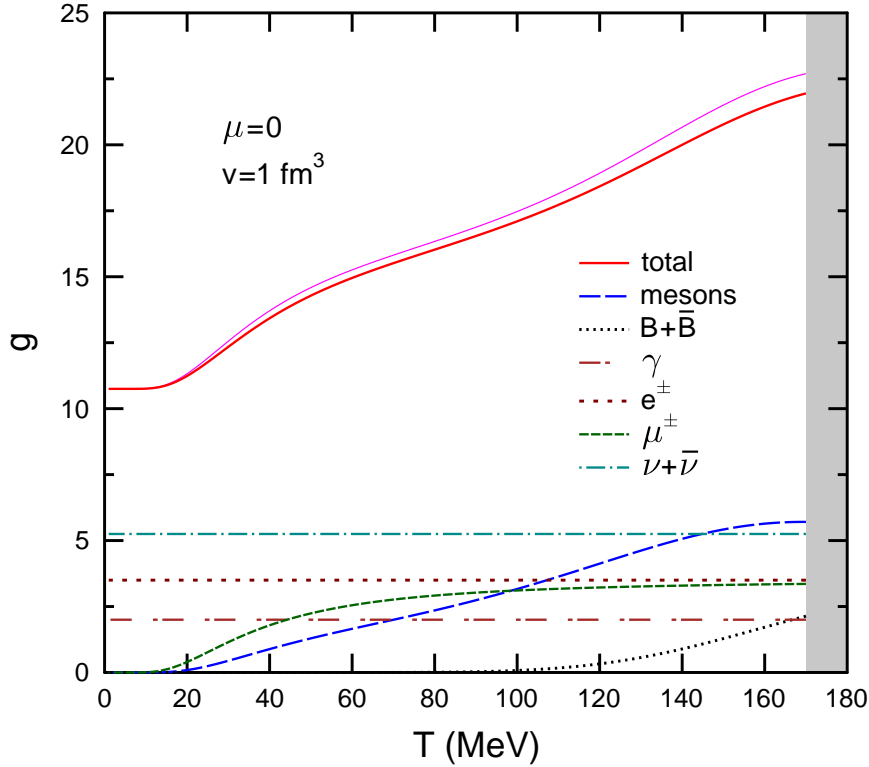}}
        \caption{(Color online)
        The effective numbers of d.o.f. (see text) as functions of temperature in the baryon--symmetric cosmic matter.
        Shading shows expected region of deconfined phase. Thin line shows $g_\varepsilon(T)$.}
        \label{fig1}
\end{figure*}
We study the evolution of primordial antibaryon abundance proceeding from~\re{kineqa2}. In addition to $Y$, it is useful to
introduce the observable quantity, the antibaryon to photon ratio $\eta=n_{\ov{B}}/n_{\hspm\gamma}$, where $n_{\hspm\gamma}$ is the density of photons. One can use the relations for the entropy- and number densities of photons
\bel{psden}
s_{\hspm\gamma}=\frac{\ds 4\hsp\pi^2}
{\ds 45}\hsp T^{\hsp 3}=\frac{4\hsp \varepsilon_\gamma}{3\hsp T}\,,~~n_{\hspm\gamma}=\frac{\ds 2\hsp\zeta (3)}{\ds\pi^2}\hsp T^{\hsp 3}\simeq\frac{s_{\hspm\gamma}}{3.6}\,,
\ee
where $\varepsilon_\gamma$ is the energy density of photons and $\zeta (3)\simeq 1.202$\hspm.
Below we also introduce the effective numbers of degrees of freedom (d.o.f.)
\bel{efdof}
g=\frac{2\hsp s}{s_{\hspm\gamma}}\,,~~~g_\varepsilon=\frac{2\hsp\varepsilon}{\varepsilon_\gamma}\,,
\ee
where $s$ and $\varepsilon$ are the total entropy- and energy densities which include contributions of photon, leptons and hadrons.
Note, that at the ''radiation dominated'' epoch, when most important d.o.f. are ultrarelativistic particles with masses much
smaller than $T$, $g_\varepsilon\simeq g$\hspm . From Eqs.~(\ref{psden})--(\ref{efdof}) one has
\bel{psden1}
\eta=\frac{sY}{n_{\hspm\gamma}}=\frac{\pi^4gY}{45\hsp\zeta(3)}\simeq 1.8\hsp gY.
\ee

We have calculated contributions of various species to $\varepsilon,s,g$ as functions of temperature by using the hadronic EoS
described in Ref.~\cite{Sat09}. Figure~\ref{fig1} shows the total number of d.o.f., \mbox{$g\hsp (T)\propto s/T^3$}, as well as contributions from mesons, (anti)baryons, photons, and leptons. We take into account $e$\hsp\hsp -, $\mu$\hsp - and $\tau$-neutrinos and antineutrinos which are considered to be massless. One can see that the contribution of hadrons to entropy is about 40\% at $T\sim 17\hspm 0$~MeV
and remains noticeable down to temperatures $T\sim 40~\textrm{MeV}$. This contribution is mostly due to pions and other mesons.
The baryon-antibaryon pairs become relatively important only at $T\gtrsim 12\hspm 0~\textrm{MeV}$.
However, even at such temperatures they contribute no more than 10\% of the total entropy. It is interesting to note
that excluded volume corrections suppress significantly the hadronic parts of entropy- and energy densities.
At $T\sim 15\hspm 0~\textrm{MeV}$ the reduction factor as compared to the ideal gas is about 1/2~\cite{Sat09}.

For comparison, the thin line in Fig.~\ref{fig1} shows the result of the $g_\varepsilon$--calculation. One can see that $g_\varepsilon$ exceeds $g$ by no more than 3\% in the considered temperature interval. At~$T\lesssim 20~\textrm{MeV}$ both quantities
practically coincide. In this domain $g$ approximately equals its asymptotic (radiation dominated) value
$g_{\hsp\rm as}=g_\gamma+g_{e^\pm}+g_\nu+g_{\ov{\nu}}=10.75$~\cite{Kol90}. As we shall see below, practically all primordial antibaryons
of the Universe disappear at this stage. Therefore, the presently observed ratio $(B/\gamma)_{\hsp\rm obs}$ can be estimated from
the $n_{\rm net}/n_{\hspm\gamma}$ ratio at~$t\to\infty$. By using the same arguments as in deriving~\re{psden1} one may write the relation
\bel{obbg}
\left(\frac{B}{\gamma}\right)_{\hspace*{-1pt}\rm obs}\simeq 1.8\hsp g_{\hsp\rm as}Y_0\frac{4}{11}\,.
\ee
Here the last term takes into account that due to the $e^+e^-$ annihilations at $T\lesssim 0.5~\textrm{MeV}$, the number of photons
increases roughly by the factor $1+g_{e^+e^-}/g_\gamma\simeq 11/4$\hsp. Substituting $(B/\gamma)_{\hsp\rm obs}\simeq  6.2\cdot 10^{-10}$~\cite{Ber12} we get the estimate $Y_0\simeq 8.8\cdot 10^{-11}$.

In principle, to solve numerically the differential equation~(\ref{kineqa2}), one should know the time dependence of temperature.
But it is clear that the temperature dependence of the cooling rate, $|\dot{T}|$, is sufficient in our case. As will be shown below, the latter is inversely proportional to the characteristic expansion time\hsp\footnote
{
Note, that $\tau_{\rm exp}=(3H)^{-1}$ where $H$ is the Hubble parameter.
},
$\tau_{\rm exp}\equiv V/\dot{V}$. By using the Friedmann equation (for a flat Universe)~\cite{Kol90} one has
\bel{freq}
\tau_{\rm exp}=\frac{R}{3\dot{R}}=(24\pi G\varepsilon)^{-1/2}\simeq \frac{0.201}{\sqrt{g_\varepsilon}}\frac{M_P}{T^2}\hsp,
\ee
where $G$ is the Newton gravitational constant and \mbox{$M_P=G^{-1/2}\simeq 1.22\cdot 10^{19}~\textrm{GeV}$} is the Plank mass.

     \begin{figure*}[hbt!]
          \centerline{\includegraphics[trim=0 8.5cm 0 8cm, clip, width=\textwidth]{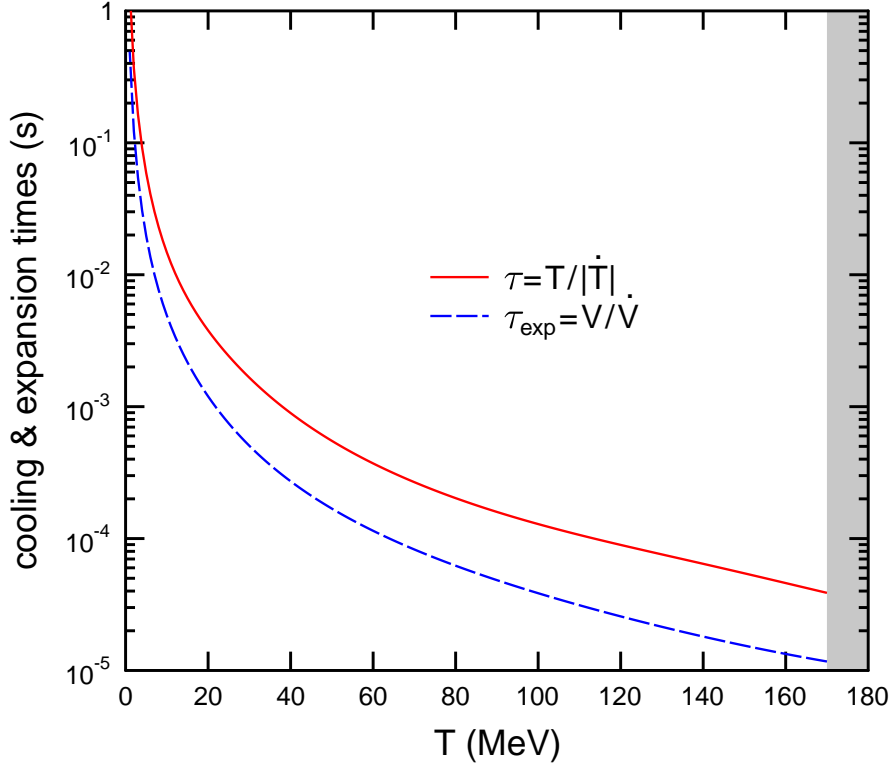}}
        \caption{(Color online)
         The expansion (the dashed line) and cooling (the solid line) times as functions of temperature
	     in the baryon-symmetric Universe.  Shading shows the region of deconfined matter.}
        \label{fig2}
        \end{figure*}
From~\re{entrc} we get the relations $\tau_{\rm exp}=\frac{\ds s}{\ds |\dot{s}|}=\frac{\ds s}{\ds |\dot{T}|s^{\,\prime}}$\,. Introducing further the charac\-teristic cooling time, $\tau\equiv\frac{\ds T}{\ds |\dot{T}|}$\,, one has
\bel{cltp}
\tau=\frac{\tau_{\rm exp}}{c_s^{\hsp 2}}=\tau_{\rm exp}\left(3+\frac{Tg^{\,\prime}}{g}\right),
\ee
were $c_s^{\hsp 2}=\frac{\ds s}{\ds T\hspace*{-1pt}s^{\,\prime}}$ is the adiabatic sound velocity squared. Figure~\ref{fig2} shows the temperature dependence  of $\tau_{\rm exp}$ and $\tau$ calculated for realistic $g\hsp (T)$ from Fig.~\ref{fig1}. One can see that at given temperature the cooling time exceeds $\tau_{\rm exp}$ by a factor of about~3. As compared to heavy--ion collisions (see next section) where $\tau_{\exp}$ is of the order of several fm/$c$, much larger values $\tau_{\rm exp}\gtrsim 10^{-5}\,s$ are characteristic for the early Universe. This difference follows from relative weakness of the gravitational interaction in combination with a spatially homogeneous character of cosmic expansion.

By using the relations $\dot{Y}=\dot{T}\,Y^{\prime}=-T\hsp Y^{\prime}/\tau$ one can rewrite~\re{kineqa2} in the form
\bel{kineqa3}
T\,\frac{dY}{d\hsp T}=\Lambda\left[\hsp Y\hsp (Y+Y_0)-Y_{\rm eq}\hsp (Y_0+Y_{\rm eq})\hsp\right],
\ee
where $\Lambda$ is a dimensionless parameter
\bel{parla}
\Lambda=\Gamma\hsp\tau=c_{\hsp V}<\sigma_{\rm ann}\hsp v_{\hspm\rm rel}>\tau_{\rm exp}\,.
\ee
In the second equality of~\re{parla} we have introduced the heat capacity per unit volume
\mbox{$c_{\hsp V}=T\hspace*{-1pt}s^{\hsp\prime}=c_s^{-2}s$}\,.
Note, that a similar form of Eqs.~(\ref{kineqa3})--(\ref{parla}) has been obtained earlier in~\cite{Sch09}.
But in contrast to our approach, the authors of Ref.~\cite{Sch09} have included only nucleons in the baryonic sector\hsp\footnote
{
In particular, annihilation of antinucleons on hyperons and baryon resonances has been neglected.
}.
The parameter $\Lambda$ changes with temperature roughly as~$\sqrt{g}\,T$. The calculation shows that $\Lambda$ increases from about \mbox{$10^{18}$} to \mbox{$5\cdot 10^{\hsp 20}$} in the temperature interval from 1 to 17\hspm 0 MeV.

     \begin{figure*}[hbt!]
          \centerline{\includegraphics[trim=0 8.5cm 0 8cm, clip, width=\textwidth]{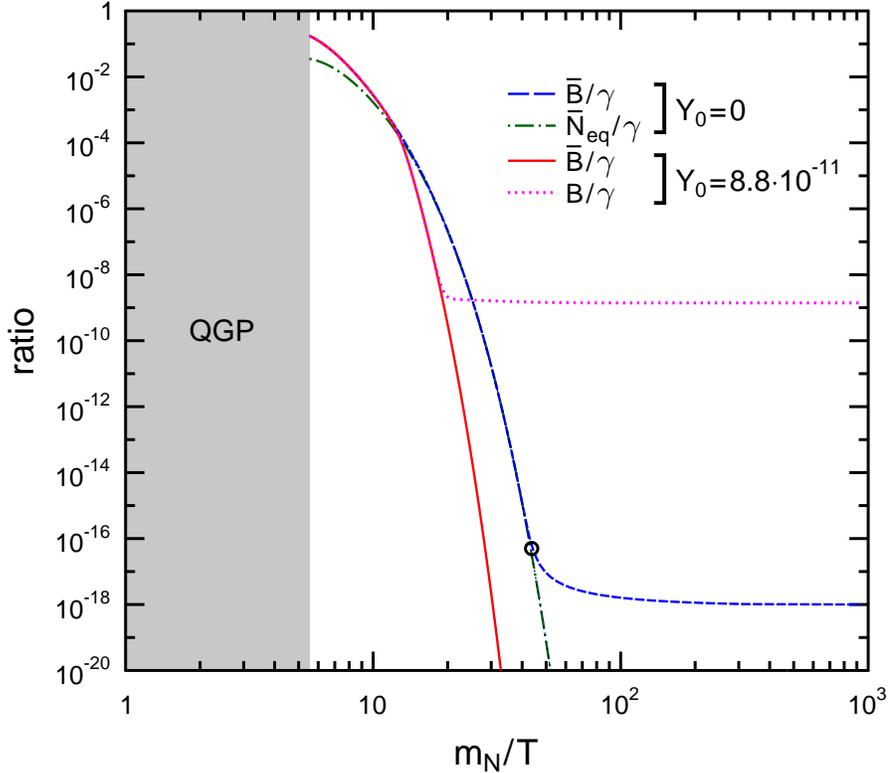}}
        \caption{(Color online)
        (Anti)baryon to photon ratios in the early Universe as functions of inverse temperature (normalized to $m_N$) are shown for different values of baryon asymmetry parameter~$Y_{\hsp 0}$. The dash-dotted line shows the equilibrium $\ov{N}/\gamma$ ratio in the baryon-symmetric case. The dot marks the freeze-out point for $Y_{\hsp 0}=0$. Shading shows the region of deconfined matter.}
        \label{fig3}
        \end{figure*}
\begin{figure*}[hbt!]
          \centerline{\includegraphics[trim=0 8.5cm 0 8cm, clip, width=\textwidth]{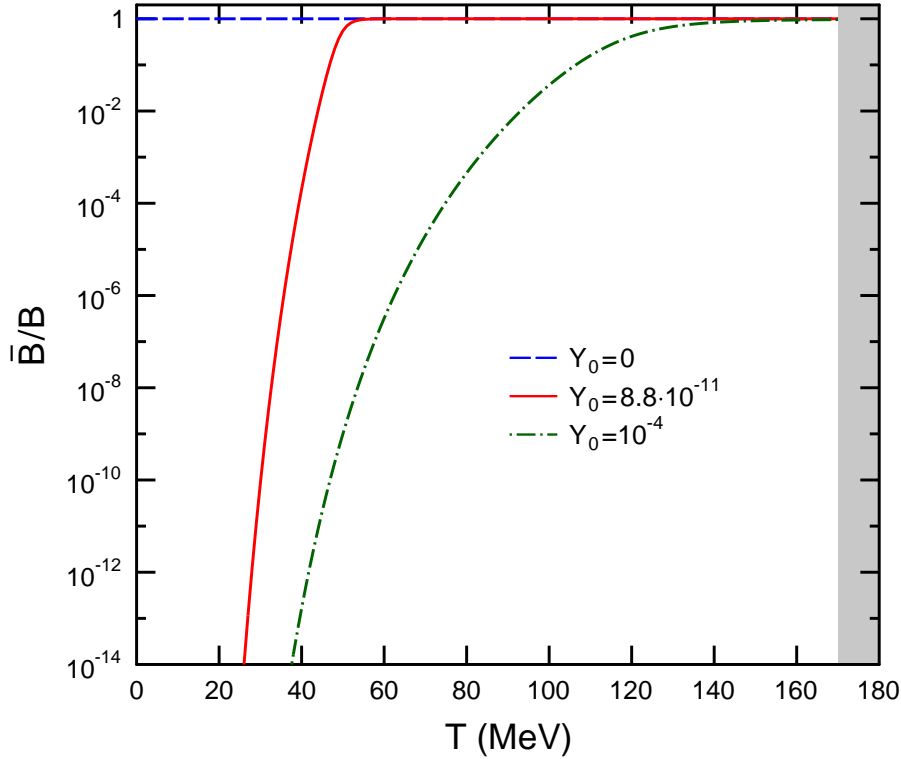}}
        \caption{(Color online)
        Same as Fig.~\ref{fig3}, but for antibaryon--to--baryon ratios as functions of temperature.}
        \label{fig4}
        \end{figure*}

Below we solve numerically the rate equation~(\ref{kineqa3}) assuming that at the initial stage $T=T_0$ the deviation from
chemical equilibrium is small, i.e. $Y(T_0)=Y_{\rm eq}(T_0)$\hsp.
We choose $T_0=17\hspm 0~\textrm{MeV}$, close to the temperature value predicted by the lattice calculations~\cite{Kar01} for the deconfinement crossover transition at vanishing chemical potential. Instead of $Y(T)$ we show in Fig.~\ref{fig3} the antibaryon-to-photon ratio $\eta$ (see~\re{psden1}) as a function of $x$. The solid curve corresponds to the baryon asymmetry parameter $Y_{\hsp 0}$ estimated from the currently observed $B/\gamma$ ratio.
The dotted curve represents the temperature dependence of the baryon-to-photon ratio $B/\gamma=1.8\hsp g\hsp (Y+Y_0)$. One can see that both lines practically coincide at $T\gtrsim 50~\textrm{MeV}$ (i.e. at $x\lesssim 20$). At lower temperatures (larger $x$) the relative fraction of antibaryons,  $\ov{B}/B\simeq Y_{\rm eq}/Y_0$, drops exponentially and rapidly becomes extremely small (see Fig.~\ref{fig4}). At temperatures between 1 and 50 MeV, $B/\gamma\simeq\frac{11}{4}\hsp (B/\gamma)_{\hsp\rm obs}\simeq 1.7\cdot 10^{-9}$.

The dashed line in Fig.~\ref{fig3} shows the results for the baryon-symmetric case~\mbox{$Y_0=0$}\hsp . One can see that at large enough temperatures the antibaryon-to-photon ratio only weakly depends on the asymmetry parameter $Y_0$\hspm . For comparison,
at the same plot we show the equilibrium antinucleon-to-photon ratio. By comparing it with the $\ov{B}/\gamma$ line, one can conclude
that excitation of antihyperons and antibaryon resonances is important only at~$T\gtrsim 100~\textrm{MeV}$.

The concept of chemical freeze-out is often used in the literature (see e.g.~\cite{Zel71,Sch86,Kol90}) to characterize deviation
of particle abundances from their equilibrium values. We postulate the system is at chemical freeze-out when the deviation from equilibrium, \mbox{$\Delta=Y-Y_{\rm\hsp eq}$}\,, satisfies the condition $\Delta\gtrsim Y_{\rm\hsp eq}$\hsp . Our calculations show that in the baryon-symmetric case this happens at $x\gtrsim x_F\simeq 45$ which corresponds to temperatures below 20~MeV.
According to Fig.~\ref{fig3}, at such temperatures the calculated values of $\eta$ (the dashed line) noticeably exceed equilibrium (anti)nucleon-to-photon ratios (the dashed-dotted line). At nonzero asymmetry parameter $Y_0\sim 9\cdot 10^{-11}$, the freeze-out for antibaryons occurs at much lower temperatures \mbox{$T\lesssim 6~\textrm{MeV}$}. This correspond to extremely small antibaryon-to-photon ratios~\mbox{$\eta\lesssim 10^{-73}$}.

Following~\cite{Zel71,Sch86} one can get analytic estimates by using~\re{kineqa3} with  $Y_{\hsp 0}=0$\hspm .
In the vicinity of the freeze-out point one has
 \bel{fest}
 \Delta\simeq\frac{\ds x}{\ds 2\hsp\Lambda Y_{\rm eq}}\left|\frac{\ds dY_{\rm eq}}{\ds dx}\right|\simeq Y_{\rm eq}\,.
 \ee
Substituting the equilibrium ratio\hsp\footnote
{
In calculating the equilibrium antinucleon density, $n_{\ov{N}}^{\rm (eq)}$, we neglect the quantum degeneracy effects.
}
\bel{yeqn}
Y_{\rm eq}\simeq\frac{n_{\ov{N}}^{\rm (eq)}}{s}\simeq\frac{45\, x^2}{\pi^4 g}\hsp K_2(x)\,,
\ee
where $K_n$ is the MacDonald function of the $n$-th order, leads to
\bel{fest1}
\frac{K_1(\hspm x)}{2K_2^{\hsp 2}(x)}\simeq\frac{45\, x}{\pi^4 g}\hsp\Lambda\equiv\lambda (x)\hsp.
\ee
In the limit $x\gg 1$ one can replace $\lambda$ by its asymptotic value
\mbox{$\lambda_{\hsp\infty}=\lim\limits_{x\to\infty}\lambda\simeq 4\cdot 10^{19}$}.
Taking into account large values of $\lambda$, one can write down an approximate solution of~\re{fest1}\hsp , $x=x_F$, in the form
\bel{fest2}
x_F\simeq \ln\left(\lambda_{\hsp\infty}\sqrt{\frac{2\pi}{\ln{\lambda_{\hsp\infty}}}}\right)\simeq 44\,.
\ee
This value agrees well with our numerical calculation.
Omitting the production term in ~\re{kineqa2}, one can estimate the asymptotic value of~$\eta$ as
\mbox{$\eta_{\hsp\infty}\simeq \frac{\ds x_F}{\ds\zeta (3)\lambda_{\hsp\infty}}\simeq 8.8\cdot 10^{-19}$}. This value
overestimates the ''exact'' $\eta\hsp (t\to\infty)$ value only by 3\%.
\vspace*{3mm}

Figure~\ref{fig4} shows the ratios $n_{\ov{B}}/n_B=Y/(Y+Y_0)$ as functions of temperature for several values of the
parameter $Y_0$\hspm . The choice $Y_0=10^{-4}$ roughly corresponds to the net baryon-to-entropy ratio in Pb+Pb collisions at
the LHC bombarding energy (see next section). In this case noticeable deviations of the $\ov{B}/B$ ratio from unity occurs
already at $T\lesssim 12\hspm 0~\textrm{MeV}$.

\section{Antibaryons in nuclear collisions}

In this section we consider the evolution of (anti)baryon abundances in relativistic heavy--ion collisions.
We focus mainly at most central Pb+Pb collisions at the SPS (\mbox{$E_{\hsp\rm lab}=158~\textrm{AGeV}$}) and LHC ($\sqrt{s_{NN}}=2.76~\textrm{TeV}$) bombarding energies. Also, central Au+Au collisions at the RHIC energy $\sqrt{s_{NN}}=200~\textrm{GeV}$ will be discussed. For recent reviews of experimental and theoretical results concerning these reactions, see Refs.~\cite{Fri11,Mue12}.

As compared to the early Universe, the dynamics of matter created in heavy--ion collisions is essentially more complicated.
This follows from much larger spatial gradients and expansion rates of multiparticle systems produced in such processes.
As a consequence, deviations from local thermodynamic equilibrium should be rather important at least at late stages
of a nuclear collision. An additional difficulty appears due to a very complicated and poorly known hadronization dynamics of
rapidly expanding QGP which is believed to be formed in such collisions.

On the other hand, the formation of photons and leptons is apparently not so important for global dynamics of hadronic
systems produced in nuclear collisions. Indeed, typical sizes of such systems are much smaller than mean free paths
of electromagnetically and weakly interacting particles. We assume that these particles escape freely into vacuum
and do not change significantly the entropy and energy of hadronic matter produced in such collisions.

For our qualitative analysis we assume that a locally equilibrated, spatially homogeneous system of hadrons (''fireball'')
is formed at some intermediate stage ($t=t_0$) of a heavy--ion collision. Below we are mainly interested in the evolution of particle densities in a ''central slice'' of the system, which corresponds to space-time rapidities $|\eta|=\tanh^{-1}|z|/t\lesssim 1$ in the c.m. frame (here $z=0$ corresponds to the symmetry plane transversal to the beam axis $z$)\hsp\footnote
{
The transfer of entropy and baryon charge from this slice~\cite{Sat07} is neglected.\label{ftn1}
}.
Disregarding dissipation effects  we again assume
the isentropic character of system expansion i.e. postulate that~$\sigma\simeq\textrm{const}$ at \mbox{$t>t_0$}\hsp .
Note, that now the entropy density includes the contribution of hadrons only.

\renewcommand{\baselinestretch}{0.8}
\squeezetable
\begin{table}[h!]
\caption{
Temperature, chemical potentials and entropy per net baryon obtained from thermal
fits of hadron ratios in central AuAu and PbPb
collisions at different c.m. bombarding energy $\sqrt{s_{NN}}$
}
\label{tab1}
\bigskip
\begin{ruledtabular}
\begin{tabular}{c|c|c|c}
\footnotesize
$\sqrt{s_{NN}}$ \hspm (GeV)  & 17.3 & 200   & 2760\\[1pt]
\hline
$T$      \hspm (MeV)         & 159  & 164   & 164 \\
$\mu$    \hspm (MeV)         & 219  & 22    & 1.7 \\
$\mu_S$  \hspm (MeV)         & 49.5 & 4.9   & 0.38 \\
$\sigma$                     & 35.7 & 368   & 4769 \\
\end{tabular}
\end{ruledtabular}
\normalsize
\vspace*{-3mm}
\end{table}
\renewcommand{\baselinestretch}{1.2}
In Table~\ref{tab1} we present the ''freeze-out values'' of temperature $T$, baryon ($\mu$) and strange~($\mu_S$) chemical potentials, as well as the entropy per net baryon $\sigma$, determined from thermal fits of hadron midrapidity ratios
in central heavy--ion collisions at different bombarding energies (for details, see Refs.~\cite{Sat09,And09}). Unless stated otherwise,
we use the values of $\sigma$ from this table to determine the temperature dependence of chemical potentials and equilibrium hadronic densities at given $\sqrt{s}$. One should bear in mind, that within such an approach, the midrapidity ratios of $\pi,K,\ov{K}$ mesons observed in the above-mentioned reactions are well reproduced.

We study the evolution of (anti)baryon abundances in heavy--ion collisions by using the numerical solution of~\re{kineqa2}\hsp\footnote
{
In principle, one could perform a more consistent study by using a chemically
non-equilibrium hydrodynamics with hadrochemical reactions, as proposed in Ref.~\cite{Bir83}.
}.
 As in Sec.~II, we choose the initial temperature \mbox{$T_0=17\hspm 0~\text{MeV}$} and apply the condition $Y(t_0)=Y_{\rm eq}(t_0)$\hspm , i.e. we neglect deviations from chemical equilibrium in the initial fireball. However, now we find the time dependence of temperature from~\re{entrc} by assuming a certain law of the fireball expansion, \mbox{$V=V(t)$}\hsp , consistent with hydrodynamical simulations. Two scenarios are considered: 1)~the Bjorken-like 1D--expansion along the beam axis~\cite{Bjo83}\hspm , 2)~the~3D cylindrical expansion in longitudinal as well as transverse directions.

In the Bjorken scenario one assumes a linear growth \mbox{$V\propto t$} which leads to the well--known
relation $s\hspm (t)\hsp t=s\hspm (t_0)\hsp t_0$\hspm . To simulate a cylindrical expansion, we apply the para\-me\-tri\-zation,
suggested in~Refs.~\cite{Cgr01,Nor10} (omitting acceleration in transverse directions):
\mbox{$V\propto t\left[R+v_T\hsp (t-t_0)\right]^2$}, where $R$ and $v_T$ are constant parameters. It is convenient
to represent this relation in the form
\bel{vtdep}
\frac{V}{V_0}=\frac{t}{t_0}\left(\frac{1+\alpha\, t/t_0}{1+\alpha}\right)^2,
\ee
where ~$\alpha=v_Tt_0/(R-v_Tt_0)$ is a dimensionless constant and $t>t_0$. In the limiting case~$\alpha=0$ one returns to
the 1D Bjorken expansion. In the case of central Au+Au collisions  at the RHIC energy $\sqrt{s_{NN}}=200$~GeV we choose the values $t_0=4$~fm/c~\cite{Nor10} and  $R=7$~fm. For $v_T/c$ in the range $0.3-0.6$~\cite{Mue12} we get the estimate $\alpha\sim 0.2-0.5$\,.
The quantities $\sigma,t_0,\alpha$ are essential parameters of our model which determine the (anti)baryon
abundances in heavy-ion collisions.

\begin{figure*}[hbt!]
          \centerline{\includegraphics*[trim=0 8.5cm 0 8cm, clip, width=\textwidth]{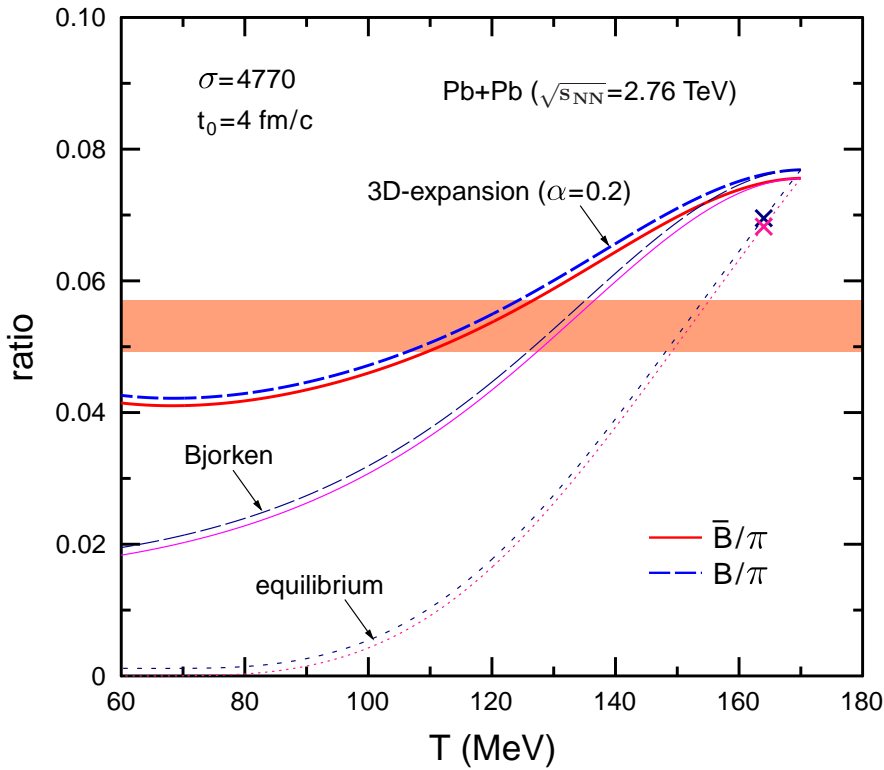}}
        \caption{(Color online)
        (Anti)baryon-to-pion ratios as functions of temperature in central Pb+Pb collisions at $\sqrt{s_{NN}}=2.76$~\textrm{TeV}. Thick solid and dashed lines show the results for 3D--expansion with the parameter \mbox{$\alpha=0.2$}\hsp. Thin lines correspond to the Bjorken scenario (\mbox{$\alpha=0$})\hspm . Dotted lines represent chemically equilibrated ratios. Shading shows experimental bounds for $\ov{B}/\pi$ ratio obtained from ALICE midrapidity data~\cite{Abe12}. Crosses mark the predictions of thermal model with parameters from Table~\ref{tab1}\hspm .}
        \label{fig5}
        \end{figure*}

\begin{figure*}[hbt!]
          \centerline{\includegraphics*[trim=0 8.5cm 0 8cm, clip, width=\textwidth]{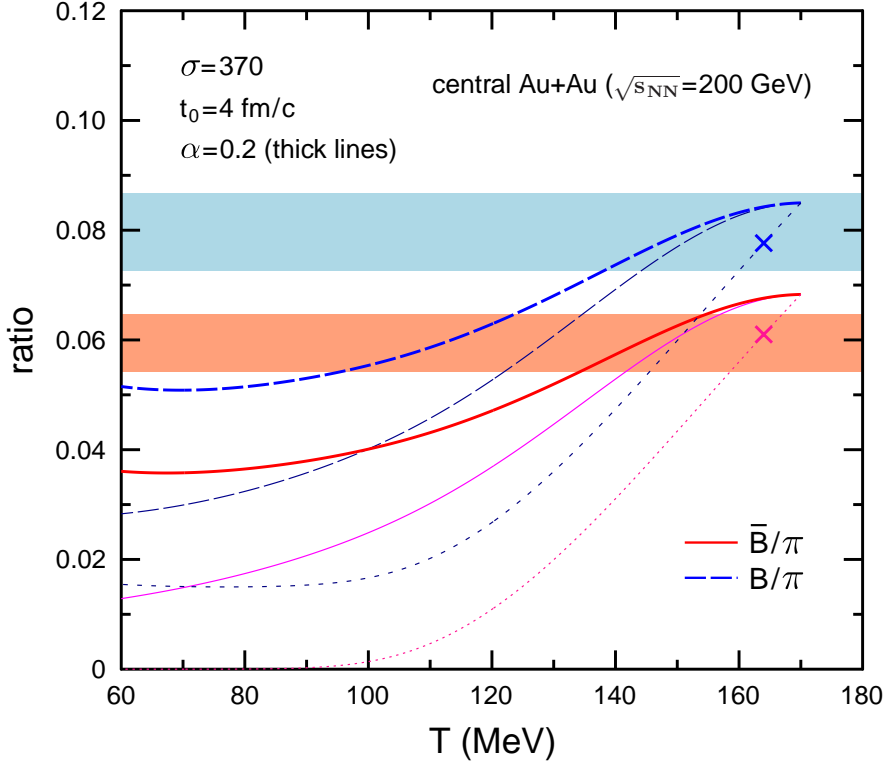}}
        \caption{(Color online)
         Same as Fig.~\ref{fig5} but for central Au+Au collisions at \mbox{$\sqrt{s_{NN}}=200$~\textrm{GeV}}.
         Shaded regions show experimental bounds for $B/\pi$ and~$\ov{B}/\pi$ ratios obtained from PHENIX~\cite{Adl04} and STAR~\cite{Ada07} data.}
        \label{fig6}
        \end{figure*}

\begin{figure*}[hbt!]
          \centerline{\includegraphics*[trim=0 8.5cm 0 8cm, clip, width=\textwidth]{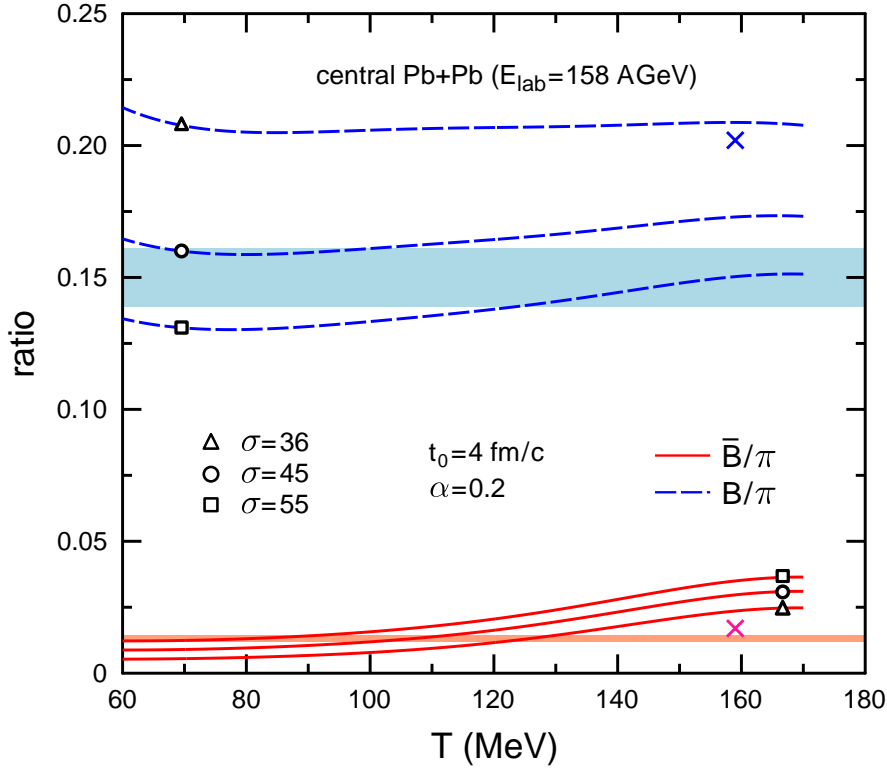}}
        \caption{(Color online)
         (Anti)baryon-to-pion ratios as functions of temperature in central Pb+Pb collisions at \mbox{$E_{\hsp\rm lab}=158$~\textrm{AGeV}}.
         Different lines correspond to different values of the parameter~$\sigma$.
         Upper and lower shaded regions show estimates of $B/\pi$ and $\ov{B}/\pi$ ratios obtained from NA49 midrapidity data~\cite{Ant11,Afa02,Ant04,Alt05,App99}. Crosses show thermal model estimates of these ratios with parameters from
         Table~\ref{tab1}\hspm .
         }
        \label{fig7}
        \end{figure*}

In Figs.~\ref{fig5}--\ref{fig7} we show the temperature dependence of (anti)baryon-to-pion ratios in central heavy--ion collisions
at LHC, RHIC and SPS energies. We use the relations
\bel{bb2p}
\frac{\ov{B}}{\pi}\equiv\frac{n_{\ov{B}}}{n_\pi^*}=Y\hsp\frac{s}{n_\pi^*}\,,
\ee
where $n_\pi^*$ is the equilibrium density of pions including those, hidden in resonances~\cite{Sat09}. The relation for $B/\pi$ is given
by the replacement \mbox{$Y\to Y+Y_0$}\hspm , where~\mbox{$Y_0=1/\sigma$}. Note, that at temperatures
$T\gtrsim m_\pi\simeq 140~\textrm{MeV}$ the $\ov{B}/\pi$ ratio is approximately proportional to the antibaryon multiplicity
in the fireball. Indeed, one can write down the total pion multiplicity as  $N_\pi=n_\pi^*V\propto n_\pi^*/s$.
Our calculation shows that the quantity $s/n_\pi^*$ decreases only by 5\%
(from 6.1 to 5.8) when the temperature decreases from 17\hspm 0 MeV to 12\hspm 0 MeV  along the adiabatic trajectories with
$\sigma\simeq 370$ (RHIC) and $\sigma\simeq 4800$~(LHC). Slightly larger variation of this
quantity (from 6.5 to 7.0 in the same temperature interval) takes place for $\sigma\simeq 40$~(SPS). \mbox{S\hspm imple} estimates show that
at later stages of the fireball expansion, corresponding to temperatures $T\lesssim 100~\textrm{MeV}$, our assumption of thermal equilibrium
is not valid anymore because the mean free paths of hadrons exceed typical fireball extensions.
\vspace*{3mm}

To make comparison with observable data possible, we estimate the asymptotic $\ov{B}/\pi$ and~$B/\pi$ ratios by using experimental
rapidity densities $dN_i/dy$ for different species \mbox{$i=\pi^{\pm}, p\hsp ,\ov{p}\hsp ,\Lambda\hsp ,\ov{\Lambda}\ldots $} at
\mbox{$y_{\hsp\rm cm}\simeq 0$}\hsp .
We use the relation
\bel{bpir}
\frac{B}{\pi}=\frac{N}{\pi}+\frac{\Lambda+\Sigma}{\pi}+\frac{\Xi}{\pi}+\frac{\Omega^-}{\pi}\,,
\ee
where \mbox{$\pi=\pi^++\pi^0+\pi^-\simeq 1.5\,(\pi^++\pi^-)$}\hsp , $N=p+n\simeq 2\hsp p$\hsp,
\mbox{$\Sigma=\Sigma^++\Sigma^{\hsp 0}+\Sigma^-\simeq 3\hsp\Sigma^{\hsp 0}$}, \mbox{$\Xi=\Xi^-+\Xi^{\hsp 0}\simeq 2\,\Xi^-$}. Approximate equalities here
are obtained assuming the isotopic symmetry of hadron production in the central rapidity region\footnote
{
Following Ref.~\cite{App99} we apply phenomenological relations $N\simeq 2.07\hsp p$\,, $\Lambda+\Sigma\simeq 1.6\,(\Lambda+\Sigma^{\hsp 0})$
at the SPS energy $E_{\hsp\rm lab}=158~\textrm{A\hsp GeV}$.
}.
Up to now, $\Sigma$ yields have not been measured in relativistic heavy--ion collision. At each bombarding energy we
find the~$\Sigma^{\hsp 0}/\Lambda$ ratio by using equilibrium ideal gas formulae with parameters from Table~\ref{tab1}\hspm . In our
estimates we take into account that observed $\Lambda$ yields include the contribution from electromagnetic decays
$\Sigma^{\hsp 0}\to\Lambda\gamma$. Similar relations are used for $\ov{B}/\pi$ with the replacement of baryons by
corresponding antibaryons.
\renewcommand{\baselinestretch}{0.8}
\squeezetable
\begin{table}[h!]
\caption{
The hadronic ratios in central heavy--ion
collisions, estimated from available experimental
data at midrapidity. Numbers in parentheses give the uncertainty
of last digit\hspm (s).
}
\label{tab2}
\bigskip
\begin{ruledtabular}
\begin{tabular}{c|l|l|l}
$\sqrt{s_{NN}}$ \hspm (GeV) &17.3 (PbPb)&200 (AuAu) &2760 (PbPb)\\[1pt]
\hline
$N/\pi$                          &0.110\hspm (8)                 &4.36\hspm (65)$\times 10^{-2}$~&3.09\hspm (32)$\times 10^{-2}$ \\
$(\Lambda+\Sigma)/\pi$           &2.36\hspm (74)$\times 10^{-2}$ &3.13\hspm (29)$\times 10^{-2}$&2.00\hspm (21)$\times 10^{-2}$  \\
$\Xi/\pi$                        &6.0\hspm (4)$\times 10^{-3}$   &4.5\hspm (6)$\times 10^{-3}$   &3.5\hspm (5)$\times 10^{-3}$   \\
$\Omega/\pi$                     &2.7\hspm (8)$\times 10^{-4}$   &2.7\hspm (4)$\times 10^{-4}$   &3.3\hspm (3)$\times 10^{-4}$   \\
\hline
$B/\pi$                           &0.150\hspm (11)                &7.96\hspm (71)$\times 10^{-2}$ &5.47\hspm (39)$\times 10^{-2}$\\
\hline
$\ov{N}/\pi$                     &7.0\hspm (6)$\times 10^{-3}$   &3.17\hspm (46)$\times 10^{-2}$ &3.00\hspm (32)$\times 10^{-2}$ \\
$(\ov{\Lambda}+\ov{\Sigma})/\pi$ &4.9\hspm (1.1)$\times 10^{-3}$ &2.37\hspm (24)$\times 10^{-2}$ &1.95\hspm (21)$\times 10^{-2}$ \\
$\ov{\Xi}/\pi$                   &1.3\hspm (1)$\times 10^{-3}$   &3.8\hspm (6)$\times 10^{-3}$   &3.3\hspm (5)$\times 10^{-3}$   \\
$\ov{\Omega}/\pi$                &1.4\hspm (6)$\times 10^{-4}$   &2.7\hspm (4)$\times 10^{-4}$   &3.3\hspm (3)$\times 10^{-4}$   \\
\hline
$\ov{B}/\pi$                     &1.33\hspm (11)$\times 10^{-2}$ &5.94\hspm (52)$\times 10^{-2}$ &5.32\hspm (39)$\times 10^{-2}$
\end{tabular}
\end{ruledtabular}
\end{table}
\renewcommand{\baselinestretch}{1.2}

To estimate (anti)baryon-to-pion ratios in central Pb+Pb collisions at the LHC energy, we use the ALICE data for midrapidity yields of
$\pi^{\pm}, p\hspm ,\ov{p}$~\cite{Abe12} and $\Xi^{\pm},\Omega^{\pm}$~\cite{Abe12b}. The ratios $(\Lambda+\Sigma)/N$ and
$(\ov{\Lambda}+\ov{\Sigma})/\ov{N}$ have been calculated within the equilibrium hadron gas model with parameters $T,\mu,\mu_S$ from Table~\ref{tab1}\hspm . In the case of central Au+Au collision at the RHIC energy $\sqrt{s_{NN}}=200~\textrm{GeV}$ we use the
PHENIX data~\cite{Adl04} to estimate the $p/\pi,\,\ov{p}/\pi$ values and the STAR data~\cite{Ada07} to find the (anti)hyperon-to-pion
ratios. The midrapidity data of the NA49 Collaboration~\cite{Ant11,Afa02,Ant04,Alt05} have been used to find experimental bounds for $\ov{B}/\pi$ and~$B/\pi$ in central Pb+Pb collisions at the SPS energy $E_{\hsp\rm lab}=158~\textrm{A\hsp GeV}$. Table~\ref{tab2} shows the
(anti)baryon-to-pion ratios for the reactions considered in this paper. The observed bounds for these ratios are marked by
horizontal stripes in Figs.~\ref{fig5}--\ref{fig7}.

Figure~\ref{fig5} presents our results for $\ov{B}/\pi$ and $B/\pi$ ratios as functions of temperature in central Pb+Pb collisions
at the LHC energy $\sqrt{s_{NN}}=2.76$~\textrm{TeV}. We use the para\-meters~$\sigma=477\hspace*{0.5pt}0$ and $t_0=4~\textrm{fm}/c$\,. The solid and dashed lines show, respectively, the $\ov{B}/\pi$ and $B/\pi$ ratios. Obviously, they are  nearly equal to each other in the limit of large $\sigma$. The shaded region in Fig.~\ref{fig5} shows experimental bounds for $\ov{B}/\pi$\hsp\footnote
{
We do not show the $B/\pi$ bounds since they practically coincide with those for $\ov{B}/\pi$ at the LHC energy~(see Table~\ref{tab2}).
}.
Thick and thin lines corresponds to different choices of the parameter $\alpha$. Our calculations show that raising $\alpha$ leads to larger deviations of (anti)baryon-to-pion ratios from their equilibrium values (the dotted curves). This follows from a more rapid decrease of temperature or, equivalently, from shorter cooling times at larger~$\alpha$\hsp . Upper and lower crosses in Fig.~\ref{fig5} correspond to the thermal model estimates of (anti)baryon-to-pion ratios at $T\simeq 165~\textrm{MeV}$ (see Table~\ref{tab1}). Indeed, one can see that this model overestimates the (anti)baryon yields observed at the LHC energy by about~25\%.

From these results we conclude that the assumption of an early saturation (chemical freeze-out) of hadron yields used in thermal models does not work, at least for (anti)baryons. On the contrary, our approach predicts a gradual decrease of (anti)baryon multiplicity up to the stage of the kinetic freeze-out at $T\lesssim 100~\textrm{MeV}$. According to Fig.~\ref{fig5}, about~40\% of initial $B\ov{B}$ pairs are annihilated to the moment when temperature drops
to~12\hspm 0~MeV.  As compared to equilibrium scenario, which is valid only at a very slow expansion, the (anti)baryon-to-pion ratios drop significantly slower with decreasing~$T$. Such a behavior can be explained by insufficient annihilation rates of $B\ov{B}$ pairs.
The role of inverse processes, in particular, multi-mesonic collisions becomes negligible at late times (see Fig.~\ref{fig9} below).

The results for central Au+Au collisions at the RHIC energy $\sqrt{s_{NN}}=200~\textrm{GeV}$ are shown in Fig.~\ref{fig6}.
In our calculation we chose the same parameters $t_0,\alpha$, but use a smaller value of specific entropy $\sigma=37\hspm 0$ (see Table I).
At this bombarding energy
baryon multiplicities noticeably exceed those for antibaryons. The thermal model predictions do not contradict the observed data in this case. Note however, that the experimental bounds in Fig.~\ref{fig6} are obtained by combining the results of two different (PHENIX and STAR) experiments. One can see, that qualitative behavior of calculated (anti)baryon-to-pion ratios is similar to that at the LHC energy.

     \begin{figure*}[hbt!]
          \centerline{\includegraphics*[trim=0 8.5cm 0 8cm, clip, width=\textwidth]{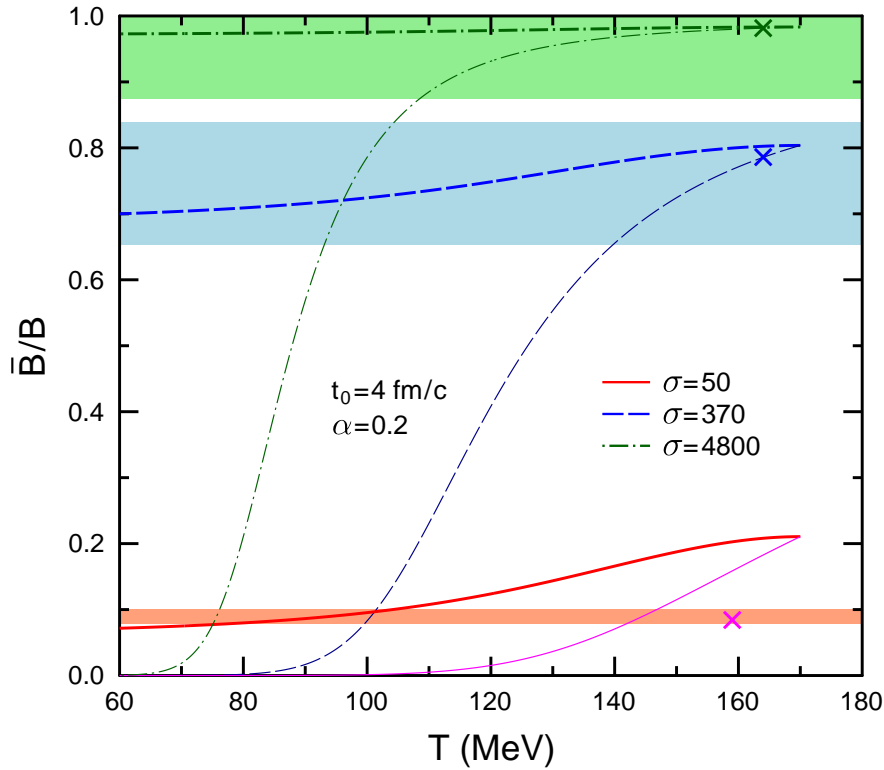}}
        \caption{(Color online)
        Antibaryon-to-baryon ratios as functions of temperature in central heavy--ion collisions for
        different values of $\sigma$.
         Upper, middle and lower shaded regions show, respectively, experimental $\ov{B}/B$ ratios obtained from ALICE (Pb+Pb, $\sqrt{s_{NN}}=2.76$~TeV), PHENIX/STAR (Au+Au, $\sqrt{s_{NN}}=200$~GeV) and NA49 (Pb+Pb, $\sqrt{s_{NN}}=17.3$~GeV) data.
         Thin lines show $\ov{B}/B$ ratios assuming chemical equilibrium.}
        \label{fig8}
        \end{figure*}
In Fig.~\ref{fig7} we present the results for central Pb+Pb collisions at the SPS incident energy $E_{\hsp\rm lab}=158~\textrm{MeV}$.
In this case the ''default'' parameter $\sigma\simeq 36$ from Table~\ref{tab1} leads to a noticeable overestimation of observed $B/\pi$ ratios. The discrepancy appears both in the thermal model and in our calculations. It is worth noting that using this $\sigma$ value leads to a significant overestimation of the $K/\pi$ ratio observed for the same reaction~\cite{Sat09}. We have checked that varying $t_0$ and $\alpha$ within reasonable limits does not remove the discrepancy with experimental $B/\pi$ ratio. We, therefore, decided to repeat calculations for different values of the parameter $\sigma$. According to Fig.~\ref{fig7}, the best agreement may be achieved for $\sigma\simeq 50$\hspm .
Note, that at the SPS energy the $\ov{B}/B$ ratio is much smaller than unity. Therefore,  $B\ov{B}$~annihilations
should not lead to a noticeable reduction of the baryon multiplicity at late stages of the reaction. But this is not true for the multiplicity of antibaryons which drops significantly, by factor of about two (see Fig.~\ref{fig8}) during the system expansion.

In Fig.~\ref{fig8} we show the $\ov{B}/B$ ratios in central Pb+Pb and Au+Au collisions at the LHC (the dashed-dotted lines),
RHIC (the dashed curves) and SPS (the solid lines) bombarding energies. For comparison, thin lines represent the equilibrium
ratios $\ov{B}/B$. Crosses again show corresponding thermal estimates with parameters from Table~\ref{tab1}\hspm .
In all three cases we take the same values of the parameters $t_0$ and $\alpha$. One can see week temperature dependencies
of $\ov{B}/B$ ratios at LHC and RHIC energies. The deviations from chemical equilibrium become stronger at lower $\sigma$.

     \begin{figure*}[hbt!]
          \centerline{\includegraphics*[trim=0 8.5cm 0 8cm, clip, width=\textwidth]{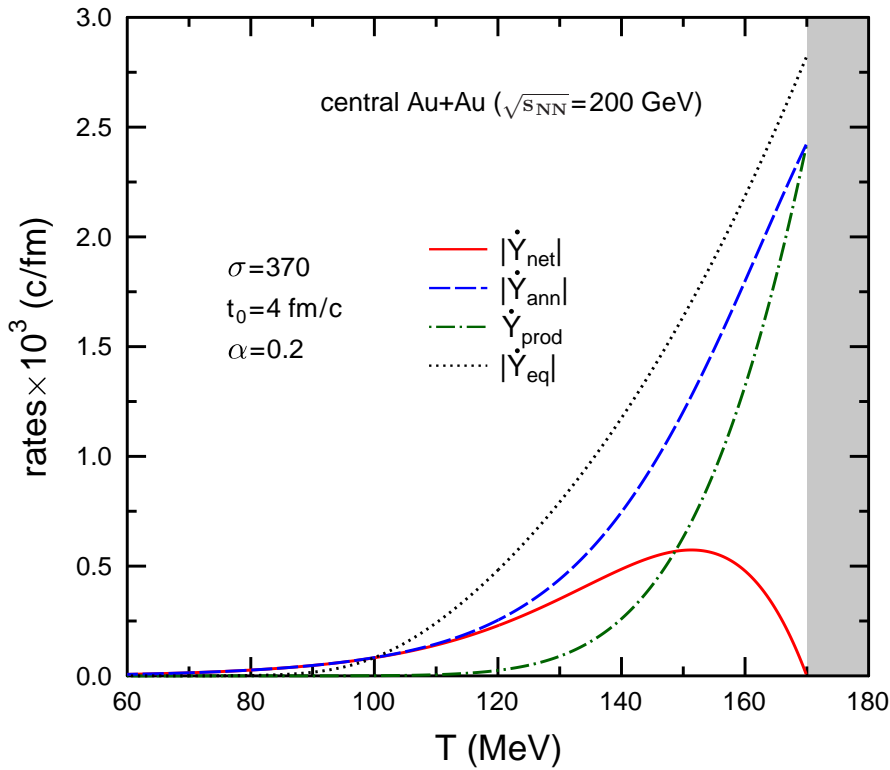}}
        \caption{(Color online)
        Time derivative of the antibaryon to entropy density ratio $Y$ in central Au+Au collision at
        $\sqrt{s_{NN}}=200$~GeV. The dashed and dotted lines show the loss (annihilation) and production components
        of $\dot{Y}$, respectively. The dotted line represents equilibrium values of $\dot{Y}$ in expanding matter.
        Shading shows the region of deconfined phase.
        }
        \label{fig9}
        \end{figure*}
It is instructive to study relative importance of the production and annihilation terms of the kinetic equation (\ref{kineqa2})
at different temperatures (they correspond, accordingly, to the first and second terms of this equation).
Figure~\ref{fig9} shows absolute values of these terms as functions of temperature in central Au+Au collisions at RHIC
energy $\sqrt{s_{NN}}=200$~GeV)\,. We choose the same values of model parameters as in Fig.~\ref{fig6}. The solid line shows
the net rate~$|\dot{Y}|$. One can see that the production and loss rates nearly compensate each other at the initial stage
of hadronic evolution corresponding to temperatures $T\gtrsim 160~\textrm{MeV}$. The calculation shows that at $T\lesssim 120~\textrm{MeV}$ the production rate becomes negligible, at later stages the multiplicity of antibaryons changes mostly due to the annihilation. The comparison of the net rate with~$|\dot{Y}_{\rm eq}|$ (the dotted curve) shows that the antibaryon abundance drops with time much slower than in chemical equilibrium. Similar trends are obtained for the LHC and SPS energies.

\section{Conclusions}

We have used the hadronic EoS with excluded volume corrections to calculate contributions of hadrons to the energy- and entropy densities
of the early Universe. It is shown that hadronic species are important at $T\gtrsim 50~\textrm{MeV}$ when they
almost double the effective number of d.o.f.
We have estimated contributions of heavy $B\ov{B}$ pairs ($B=\Lambda,\Sigma,\Delta\ldots$) as functions of temperature in expanding cosmic matter and found that they can not be neglected at early stages with $T\gtrsim 100~\textrm{MeV}$.

We have performed a similar analysis of the (anti)baryon evolution in hadronic fireballs produced in relativistic heavy--ion collisions.  We have shown that rapid fireball expansion leads to strong deviations from chemical equilibrium, which are especially large for heavy particles like (anti)baryons. We have demonstrated that the assumption of common chemical freeze-out, usually made in thermal models, is not valid at SPS, RHIC and LHC energies. Our calculations explain deviations of~$p/\pi$ and $\ov{p}/\pi$ ratios observed in Pb+Pb collisions at the LHC energy from thermal model predictions. We conclude that realistic calculations of $B,\ov{B}$ abundances in heavy--ion collisions should explicitly take into account both annihilation of (anti)baryons as well as their production in (multi)mesonic interactions. We predict that $B,\ov{B}$ multiplicities at midrapidity gradually decrease with time at least until the kinetic freeze-out stage.

Certainly, our calculations are rather crude for complicated hadronic systems produced in heavy--ion collisions. In particular, we disregarded the effects of spatial inhomogeneity and entropy non-conservation. We plan to make a more consistent study within a hydro-kinetic model taking into account deviations from chemical equilibrium as proposed in Ref.~\cite{Sat12}. One should bear in mind that the $B\ov{B}$ production terms can be enhanced if mesons are out of chemical equilibrium~\cite{Rap01,Cgr01}. In this case the evolution should be described with
chemically nonequilibrium EoS\hspm s\hsp .

A very interesting topic, not addressed in this paper is the formation and survival of antinuclei
($\ov{d},\ov{t},\ov{\alpha}\ldots$), both in the early Universe and in heavy--ion collisions. Apparently, deviations from chemical
equilibrium should be even more important in this case.

\begin{acknowledgments}
This work was supported by the Helmholtz International Center for
FAIR (Germany) and the grant NSH--215.2012.2 (Russia).
\end{acknowledgments}

\end{document}